\documentclass[a4paper,11pt]{article}

\usepackage{contribution}

\newcommand{\weblink}[2][]{%
    \ifthenelse{\equal{#1}{}}%
    {\textnormal{\url{#2}}}%
    {\textnormal{\href{#2}{#1}}}%
}

\newcommand{\acknowledgements}[1]{%
  \bigskip\bigskip
  \textsf{\textbf{\Large Acknowledgements}} \\[2ex]
  {#1}
  \bigskip
}

\def\beq{\begin{equation}}
\def\eeq#1{\label{#1}\end{equation}}
\def\eeqn{\end{equation}}

\def\beqa{\begin{eqnarray}}
\def\eeqa#1{\label{#1}\end{eqnarray}}
\def\eeqan{\end{eqnarray}}

\let\bar=\overbar

\def\etal{{\it et al.}}

\def\Dslash{\not{\hbox{\kern-4pt $D$}}}
\def\dslash{\not{\hbox{\kern-2pt $\del$}}}

\def\msb{{\bar{\ssstyle M \kern -1pt S}}}

\newcommand{\contribution}[7][]{%
  \clearpage
  \thispagestyle{plain}
  \ifthenelse{\equal{#1}{}}
  {\hypersetup{pdftitle={#2}}}
  {\hypersetup{pdftitle={#1}}}
  \hypersetup{pdfauthor={{#3} {#4}}}
  {\centering\normalfont\LARGE\bfseries\sffamily #2 \par\nobreak}
  \lhead{}
  \chead{%
    \textit{\footnotesize XIV International Conference on Hadron Spectroscopy
      (\weblink[\textit{hadron2011}]{http://www.hadron2011.de}), 13-17 June 2011, Munich, Germany}%
  }
  \rhead{}
  \bigskip
  \begin{center}
    {#3} {#4}\ifthenelse{\equal{#6}{}}{}{\footnote{\weblink[#6]{mailto:#6}}}
    \ifthenelse{\equal{#7}{}}{}{#7} \\
    \textit{#5}
  \end{center}
  \bigskip
}

\renewcommand{\abstract}[1]{%
  \begin{center}
    \begin{minipage}{0.85\textwidth}
      \begin{footnotesize}
        #1
      \end{footnotesize}
    \end{minipage}
  \end{center}
  \bigskip
}

\begin{document}


%
%
%
%
%
{  


%

\contribution[Central Meson Production in ALICE]  
{Central Meson Production in ALICE}  
{Rainer}{Schicker}  
{Phys. Inst. \\
  Philosophenweg 12\\
  D-69120 Heidelberg, GERMANY}  
{}  
{on behalf of the ALICE Collaboration}  
%

\abstract{%
 
The ALICE experiment at the Large Hadron Collider (LHC) at CERN consists 
of a central barrel, a muon spectrometer and of additional detectors 
for trigger and event classification purposes. The low transverse 
momentum threshold of the central barrel gives ALICE a unique opportunity to 
study the low mass sector of central production at the LHC.
I will report on first analysis results of meson production in
double gap events in proton-proton collisions at $\sqrt{s}$ = 7 TeV
and in PbPb collisions \mbox{at $\sqrt{s_{NN}}$ = 2.76 TeV.}

}

%


\section{Introduction}

The ALICE experiment consists of a central barrel and of a forward muon 
spectrometer \cite{Alice1}. 
Additional detectors for trigger purposes and for event classification exist 
outside of the central barrel. Such a geometry allows the 
investigation  of many topics of diffractive reactions at hadron colliders, 
for example the measurement of single and double diffractive dissociation 
cross sections and the study of central diffraction. The ALICE physics 
program foresees data taking in pp and PbPb collisions at nominal 
luminosities \mbox{$\cal{L}$ = $5\times 10^{30}cm^{-2}s^{-1}$} and 
\mbox{$\cal{L}$ = $10^{27}cm^{-2}s^{-1}$,} respectively. 
An asymmetric system pPb will be measured soon with a first test expected
this year.

\section{The ALICE Experiment}

In the ALICE central barrel, momentum reconstruction and particle 
identification are achieved in the pseudorapidity range $-1.4 < \eta < 1.4$ 
combining the information from the Inner Tracking System (ITS) and the Time 
Projection Chamber (TPC). 
In the pseudorapidity range $ -0.9 < \eta <  0.9 $, the information from the 
Transition Radiation Detector (TRD) and the Time of Flight (TOF) system is
in addition available.
A muon spectrometer covers the range $-4.0<\eta<-2.5$.
At very forward angles, the energy flow is measured by Zero Degree 
Calorimeters (ZDC) \cite{ZDC}.
Detectors for event classification and trigger purposes are located on both 
sides of the ALICE central barrel. First, the scintillator arrays V0A and V0C 
cover the pseudorapidity range $2.8 < \eta < 5.1$ and $-3.7 < \eta < -1.7$, 
respectively. The four- and eightfold  segmentation in pseudorapidity and 
azimuth result in 32 individual counters in each array.
Second, a Forward Multiplicity Detector (FMD) based on silicon strip 
technology covers the pseudorapidity range $1.7 < \eta < 5.1$ and 
$-3.4 < \eta < -1.7$, respectively.
Third, two arrays of Cherenkov radiators T0A and T0C deliver accurate timing 
for measuring the time of collisions. Figure \ref{fig:acc} shows the 
pseudorapidity coverage of these detector systems. 
    
\begin{figure}[htb]
\begin{center}
\includegraphics[width=0.5\textwidth]{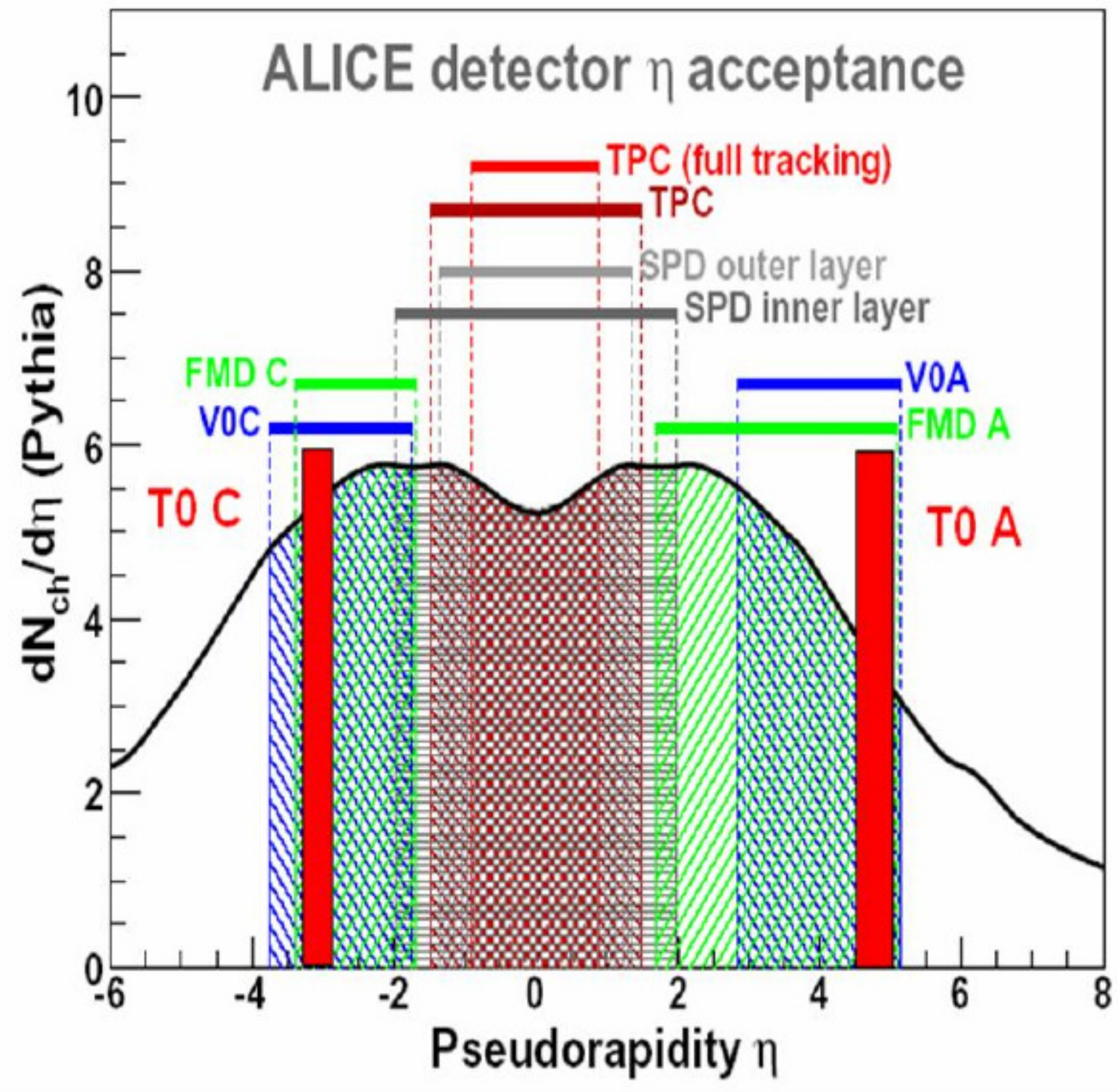}
\caption{Pseudorapidity coverage of the ALICE detectors.}
\label{fig:acc}
\end{center}
\end{figure}

\section{Central diffraction in ALICE}

Central diffractive events are experimentally defined by activity in the
central barrel and by no activity outside the central  barrel. This 
condition can be implemented in the trigger at level zero (L0) by defining
barrel activity as hits in the ITS pixel detector, or the TOF system.  
The gap condition is realized by the absence of V0 signals, hence a gap 
of two units on either barrel side can be defined at L0. In the offline 
analysis, the information from the V0, T0, FMD, SPD  and TPC detectors
define the gaps spanning the range $0.9 < \eta < 5.1$ and  
$-3.7 < \eta < -0.9$. Events with and without detector signals in these 
two ranges are defined to be no-gap and  double gap events, respectively.

A rapidity gap can be due either to Pomeron, Reggeon or photon exchange.
A double gap signature can therefore be induced by a combination of these 
exchanges. Pomeron-Pomeron events result in centrally produced states
with quantum numbers C = +1 (C = C-parity) and I = 0 (I = isospin).
The corresponding quantum numbers in photon-Pomeron induced events
are C = -1 and I = 0 or I = 1 \cite{Nachtmann}.

\section{Central meson production in pp-collisions}

In the years 2010-2011, ALICE recorded zero bias and mimimum bias data in 
pp-collisions at a center-of-mass energy of $\sqrt{s}$ = 7 TeV. 
The zero bias trigger was defined by beam bunches crossing at the ALICE
interaction point, while the minimum bias trigger was derived by 
minimum activity in either the ITS pixel or the V0 detector.
Events with double gap topology as described above are contained in this 
minimum bias trigger, hence central diffractive events were analyzed 
from the minimum bias data sample. 

For the results presented below, a sample of $3.5 \times 10^8$ minimum bias
events was analyzed.  First, the fraction of events satisfying the 
gap condition described above was calculated.  This fraction was 
found to be about $2 \times 10^{-4}$.  Only runs where  this fraction 
was calculated to be within 3 $\sigma$ of the average value of the 
corresponding distribution were further analyzed.  
This procedure resulted in about $7 \times 10^{4}$ double gap events.
As a next step, the track multiplicity in the pseudorapidity 
range $ -0.9 < \eta < 0.9$  was evaluated .

\begin{figure}[htb]
\begin{center}
\includegraphics[width=0.5\textwidth]{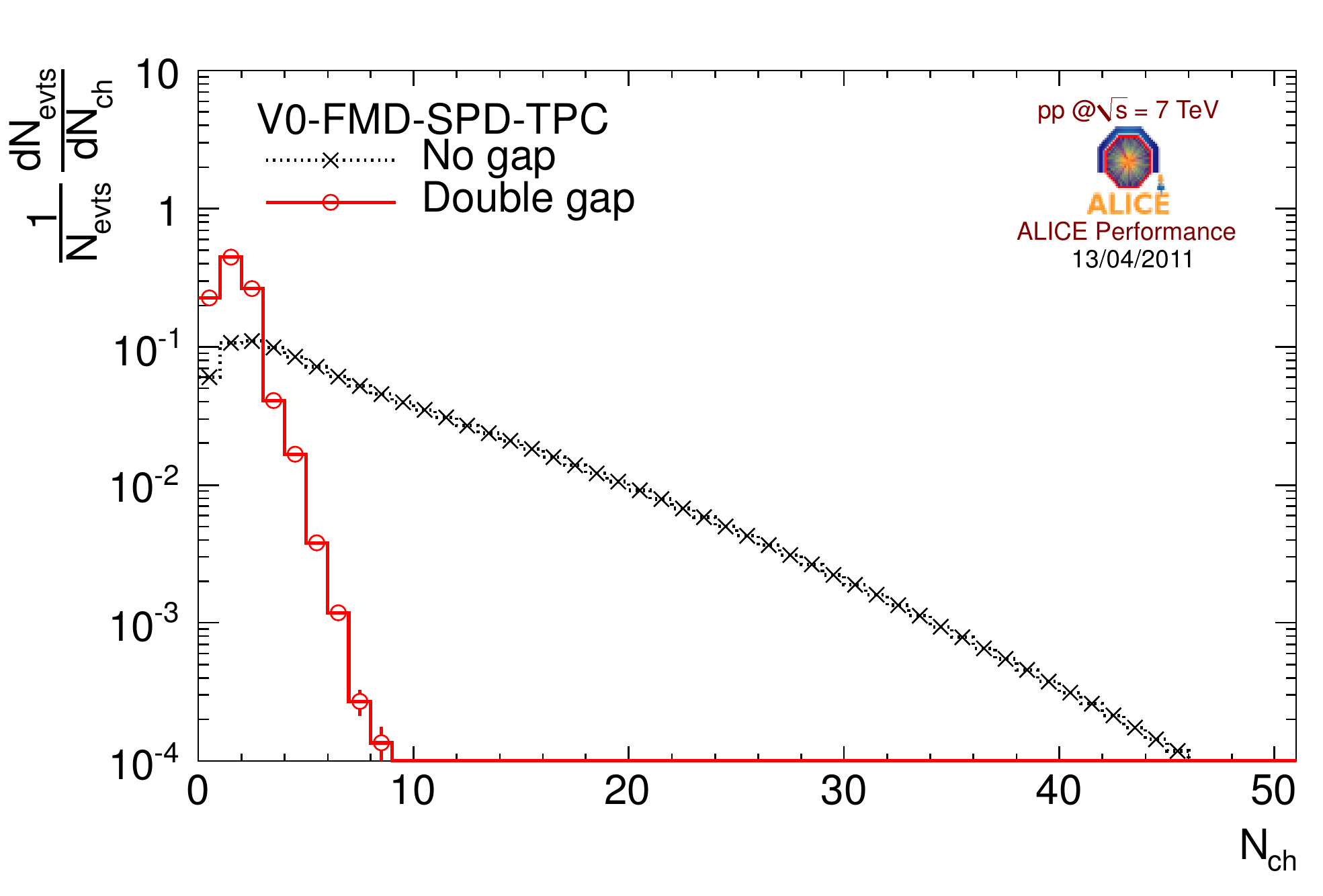}
\caption{Track multiplicity within pseudorapidity range -0.9 < $\eta$ < 0.9 for no-gap and double gap events.}
\label{fig:dgap_mult}
\end{center}
\end{figure}
   
Figure \ref{fig:dgap_mult} shows the track multiplicity in the pseudorapidity 
range \mbox{$ -0.9 < \eta < 0.9 $}  for double and 
no-gap events. Very low transverse momentum
tracks never reach  the TPC which results in events with track multiplicity 
zero. The multiplicity distributions of the double and no-gap events clearly 
show different behaviors as demonstrated  in Figure \ref{fig:dgap_mult}.

\begin{figure}[htb]
\begin{center}
\includegraphics[width=0.5\textwidth]{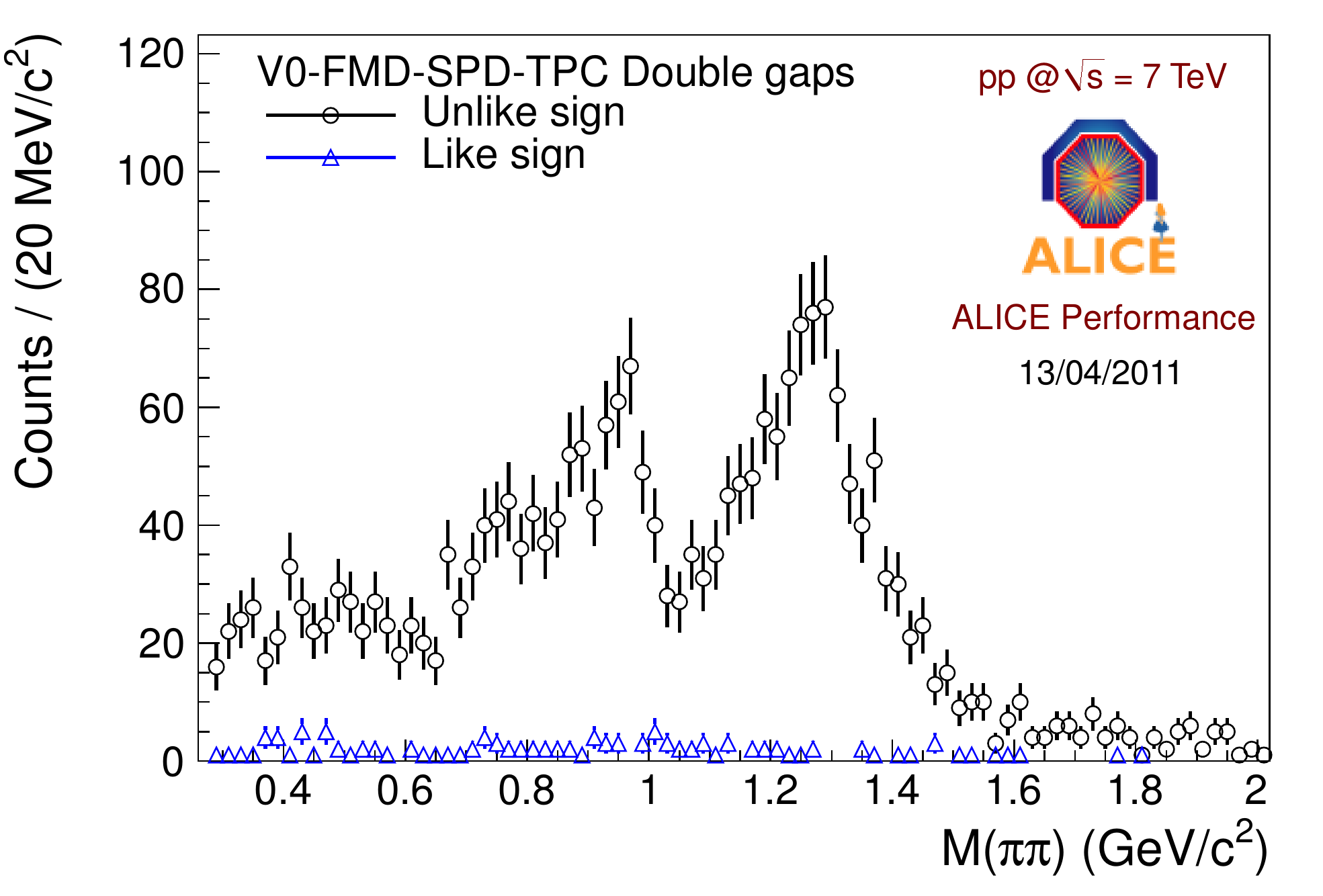}
\caption{Invariant mass distribution of like and unlike sign pion pairs.}
\label{fig:dgap_mass_lu}
\end{center}
\end{figure}

The specific energy loss dE/dx as measured by the 
TPC in combination with the TOF detector information 
identifies pions with transverse momenta p$_{T} \geq$ 300 MeV/c. 
The events with exactly two pions are selected, and the invariant mass
of the pion pairs is shown in Figure \ref{fig:dgap_mass_lu}. 
These pion pairs can be of like or unlike sign charge. Like sign pion 
pairs can arise from two pion pair production with loss of one pion 
of same charge in each pair, either due to the low p$_T$ cutoff described above,
or due to the finite pseudorapidity coverage of the detectors used for 
defining the rapidity gap. For charge symmetric detector
acceptances, the unlike sign pairs contain the signal plus background,
whereas the like sign pairs represent the background.
From the two distributions shown in Figure \ref{fig:dgap_mass_lu},
the background is estimated to be less than 5\%. 

\begin{figure}[htb]
\begin{center}
\includegraphics[width=0.5\textwidth]{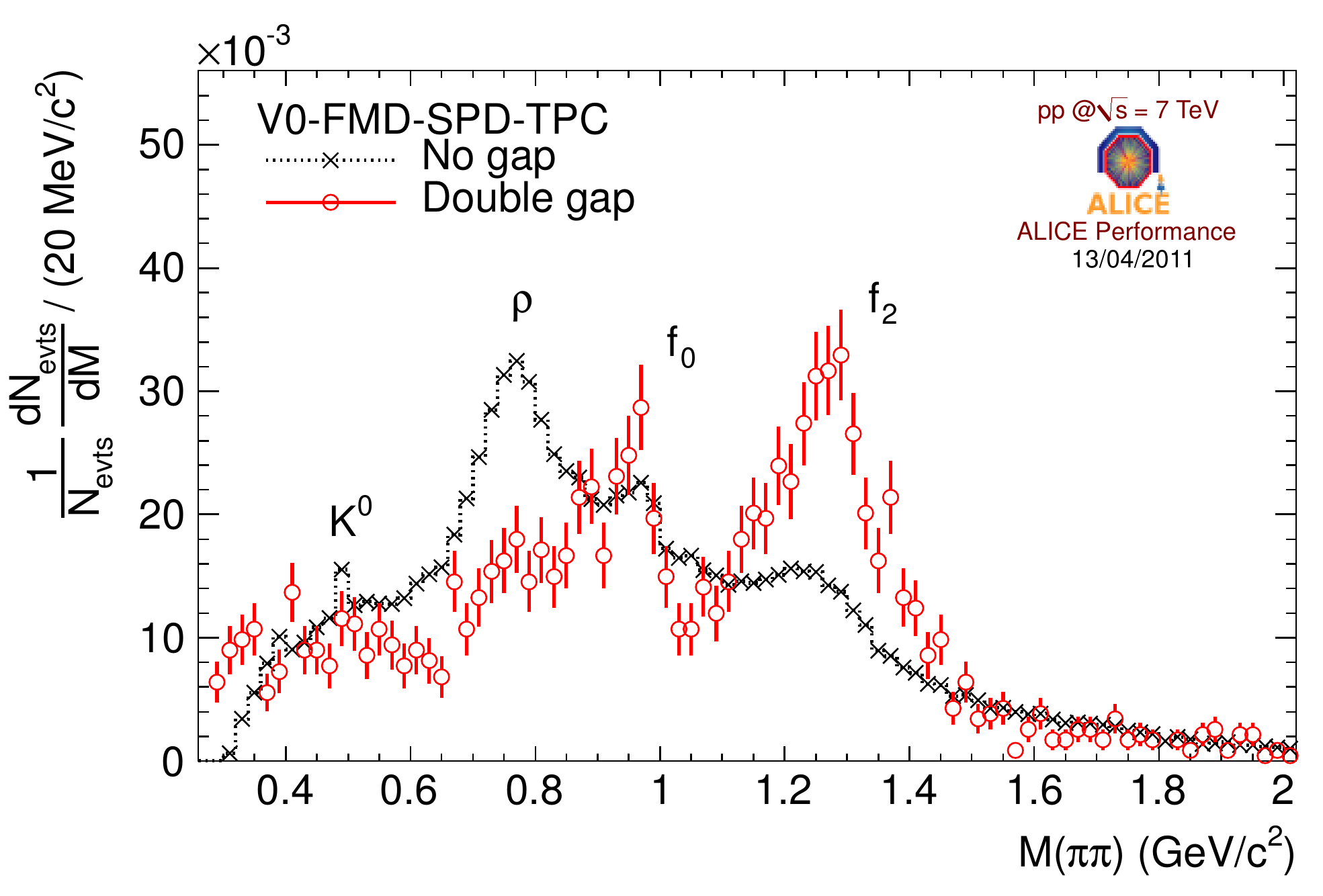}
\caption{Pion pair invariant mass distribution for double and 
for no-gap events.}
\label{fig:dgap_mass}
\end{center}
\end{figure}

Figure \ref{fig:dgap_mass} displays the normalized background corrected
pion pair mass for double and no-gap events. 
The particle identification by the TOF detector
requires the single track transverse momentum p$_T$ to be larger than 
about 300 MeV/c. This single track p$_{T}$ cut introduces a significant
acceptance reduction for pair masses $M(\pi\pi) \leq$ 0.8 $GeV/c^2$
at low pair p$_{T}$. The distributions shown are not acceptance corrected. 
In the no-gap events, structures are seen from $K^{0}_{s}$ and 
$\rho^{0}$-decays. Two additional structures are associated with f$_0$(980) 
and f$_2$(1270) decays. In the double gap distribution, the $K^{0}_{s}$ 
and $\rho^{0}$ are highly suppressed while the f$_0$(980) and f$_2$(1270) 
with quantum numbers J$^{PC}$ = (0, 2)$^{++}$ are much enhanced. This 
enhancement of \mbox{J$^{PC}$ = J$^{++}$} states is evidence that the 
double gap condition used for analysing the minimum bias data sample 
selects events dominated by double Pomeron exchange.

\section{Central meson production in PbPb-collisions}

Diffractive and electromagnetic processes in PbPb-collisions show a variety
of intriguing features \cite{Baur,Baltz}. 
First, photoabsorption can lead to giant dipole resonance excitation  
with subsequent neutron decay. In such decays, the charge to mass ratio 
is modified. In addition, bound-free pair production also leads to 
a modified charge to mass ratio of one or both of the nuclei involved. 
Both processes contribute to the beam life time.
Second, photon-photon processes result in electromagnetic production of 
pseudoscalars  $\pi^{0},\eta,\eta^{'}$ and of pairs of bosons  
$\pi^{+}\pi^{-},K^{+}K^{-}$ and  fermions 
$e^{+}e^{-},\mu^{+}\mu^{-},\tau^{+}\tau^{-}$. 
Third, photon-Pomeron processes can, for example, result in diffractive 
photoproduction of vector mesons $\rho^{0}, \phi, J/\Psi, \Upsilon$.

The first heavy ion run at the LHC took place in Nov-Dec 2010. In this 
period, about $12 \times 10^{6}$ minimum bias PbPb-collisions were 
recorded with the ALICE detectors. In addition to the minimum bias trigger, 
data were taken with two dedicated triggers
for investigating meson production in the ALICE central barrel.
First, a trigger OM2 based on number of \mbox{TOF hits $\geq 2$} was running.
Second, a trigger CCUP2 defined by  the logic combination: (TOF hits $\geq 2$) 
AND (ITS pixel) AND (double gap condition) was defined. 

\begin{figure}[htb]
\begin{center}
\includegraphics[width=0.5\textwidth]{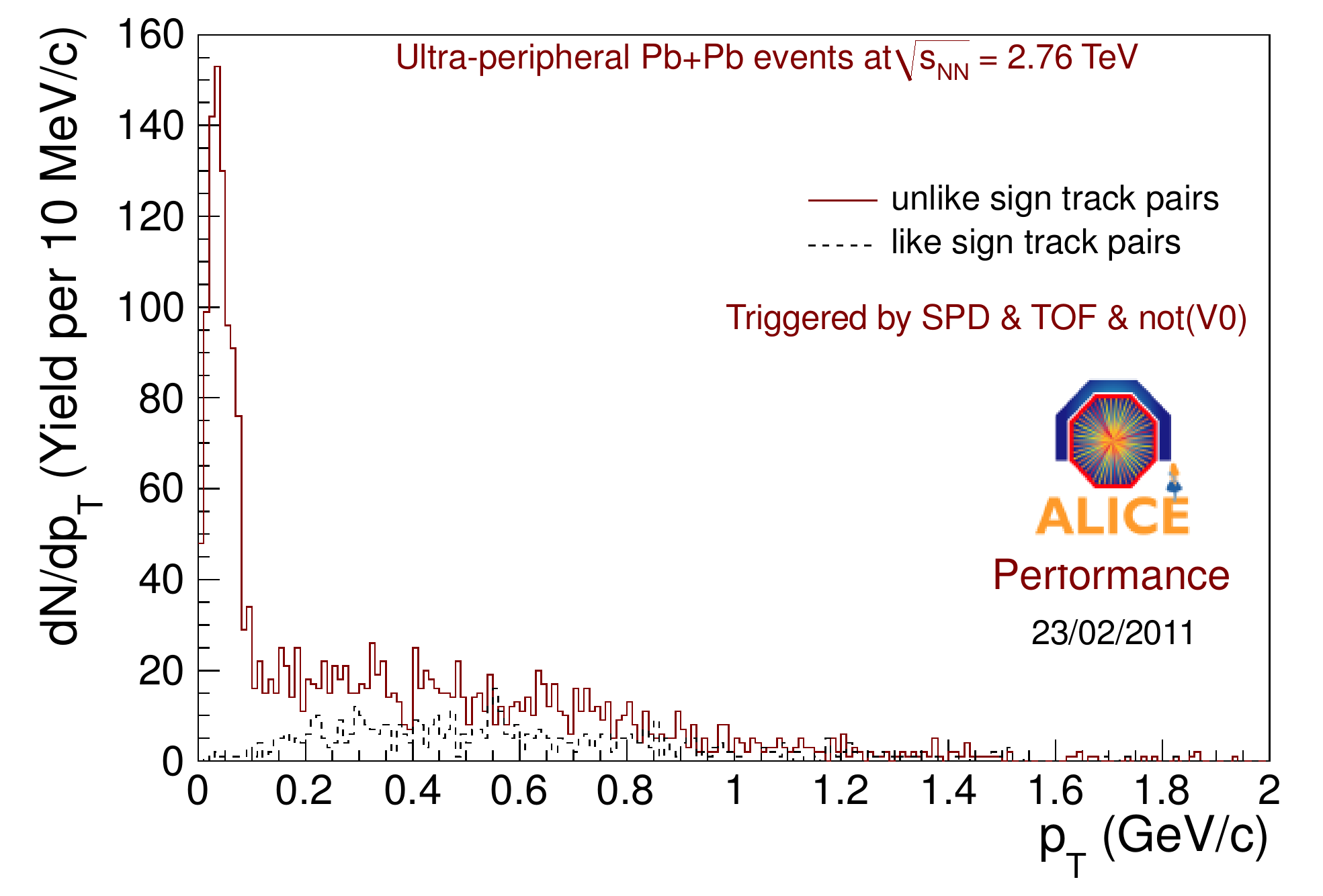}
\caption{Transverse momentum of unlike and like sign CCUP2 triggered 
pairs in PbPb-collisions at $\sqrt{s_{NN}}$ = 2.76 TeV.}
\label{fig:ccup2_pt}
\end{center}
\end{figure}

The OM2 and CCUP2 triggered  events with exactly two tracks in the central 
barrel were selected, and the pair transverse momentum was calculated.
Both of these triggers result in similar pair p$_T$ distributions.
The pair p$_T$ of CCUP2 triggered events is shown in 
\mbox{Figure \ref{fig:ccup2_pt}}  for like and  unlike sign pairs.  
The unlike sign distribution clearly shows a strong 
peak at \mbox{p$_T \leq $ 100 MeV/c} consistent with coherent
production off a Pb nucleus.

\begin{figure}[htb]
\begin{center}
\includegraphics[width=0.5\textwidth]{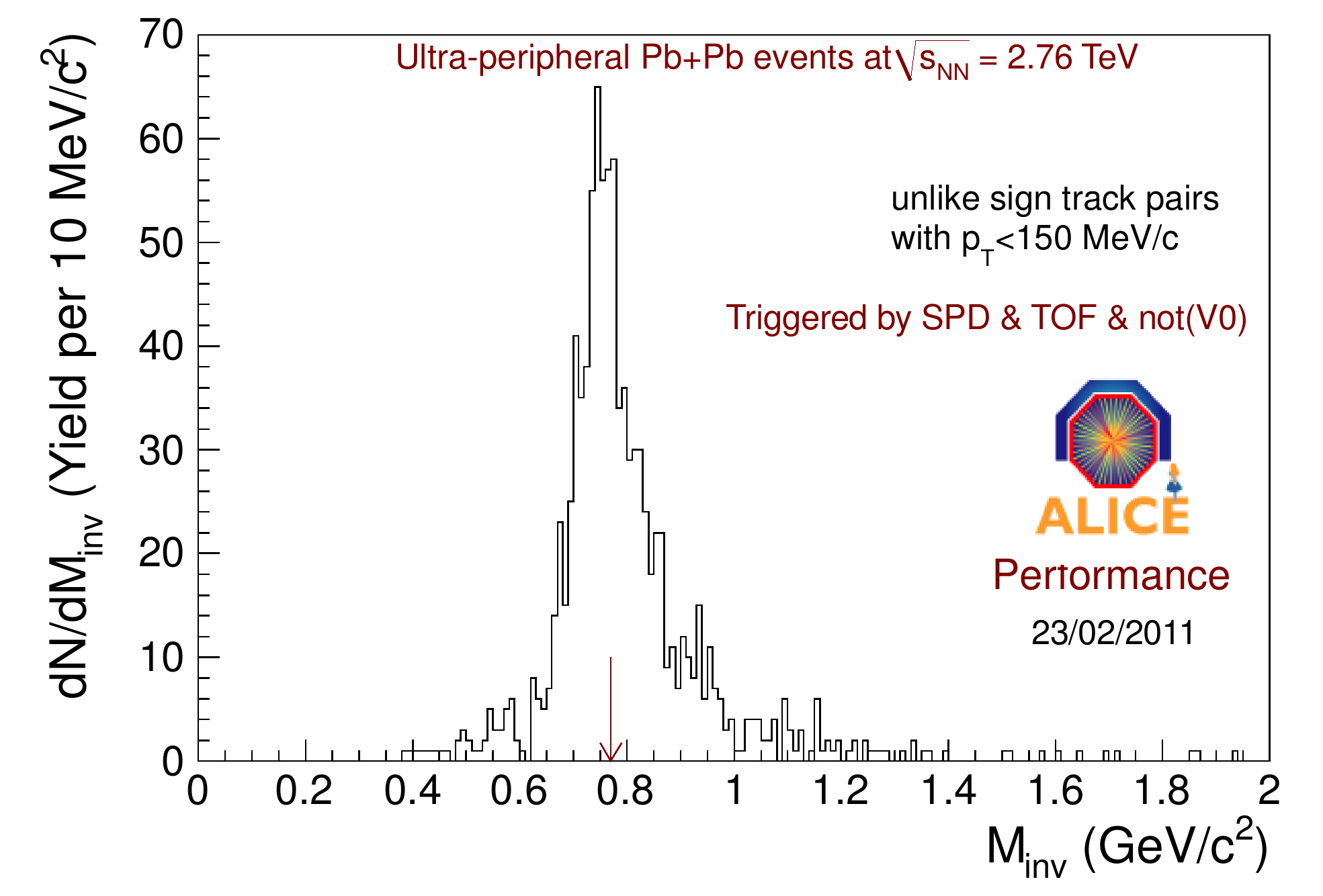}
\caption{Invariant mass of unlike sign pairs with p$_T \leq$  150 MeV/c
for CCUP2 triggered pairs in PbPb collisions at $\sqrt{s_{NN}}$ = 2.76 TeV.}
\label{fig:ccup2_mass}
\end{center}
\end{figure}

Figure \ref{fig:ccup2_mass}  shows the invariant mass for the unlike sign
CCUP2 triggered pairs with pair \mbox{p$_T \leq 150$ MeV/c.} 
This distribution is not corrected for finite detector acceptance and for 
detector resolution. The shape of the distribution shown in 
Figure \ref{fig:ccup2_mass} is consistent with production of the 
$\rho^{0}$-meson with J$^{PC}$ = 1$^{--}$. As for the p$_T$-distribution, 
the OM2 and CCUP2 triggers result in similar pair invariant mass distributions.
Hence these results indicate that double gap 
events in PbPb collisions are not dominated by 
Pomeron-Pomeron events as is the case in  proton-proton collisions. 

\section{Conclusions and Outlook}

First analysis results show that central meson production in double gap events 
at LHC energies is consistent with the hypothesis of Pomeron-Pomeron and
photon-Pomeron exchange in pp and PbPb-collisions, respectively.
The next step in the study of these events  will be a quantitative 
analysis of the results.

\acknowledgements{%
This work is supported in part by German BMBF under project 
06HD197D and by WP8 of the hadron physics program of the 
7$^{th}$  EU program period.

}


%

}  


\end{document}